\documentstyle[pre,aps,prb]{revtex}

\newcommand{\erf}{\mathop{\rm erf}\nolimits}

\begin{document}
\draft
\title{
Hydration of a B--DNA Fragment in the Method of
Atom--atom Correlation Functions with the Reference
Interaction Site Model Approximation}

\author{D.A.~Tikhonov, R.V.~Polozov,}

\address{
Institute of Theoretical and Experimental Biophysics,
Russian Academy of Sciences, Pushchino, Moscow Region, 142292, Russia}

\author{E.G.~Timoshenko
\thanks{Corresponding author; E-mail: Edward.Timoshenko@ucd.ie},
Yu.A.~Kuznetsov,  A.V.~Gorelov,  K.A.~Dawson}

\address{
Irish Centre for Colloid Science and Biomaterials,
Department of Chemistry, University College Dublin,
Belfield, Dublin 4, Ireland}

\date{\today}

\maketitle

\begin{abstract}
We propose an efficient numerical algorithm for solving integral equations
of the theory of liquids in the Reference
Interaction Site Model (RISM) approximation for infinitely dilute
solution of macromolecules with a large number of atoms. The algorithm
is based on applying the nonstationary iterative methods for solving
systems of linear algebraic equations.
We calculate the solvent--solute atom--atom
correlation functions for a fragment of the B-DNA duplex d(GGGGG)$\cdot$d(CCCCC)
in infinitely dilute aqueous solution. The obtained results are
compared with available experimental data and results from
computer simulations.
\end{abstract}


\section{Introduction}
\label{sec:int}

Interactions of nucleic acids with proteins, lipids and other ligands
in many respects are determined by the solvation properties of these
biomolecules. The structure of the DNA molecule itself in the cytozolic
environment of a cell is strongly dependent on its hydration.
Information about the structure of water shell around many biomolecules
is available primarily from X-ray and neutron scattering experiments
\cite{XrayExp} on their crystalline forms, as well as from NMR studies
\cite{NMRExp} of such molecules in solution.

There is a large number of reviews devoted to the problem of biopolymer
hydration \cite{BioHydRev}. One of common approaches to these difficult
problems is based on computer simulations using either Molecular Dynamics or
Monte Carlo methods \cite{MDMCHydra,Chuprina91,Teplukhin92}.
The main limitation of these direct
simulation techniques is in the huge computational expense for reaching
the equilibrium state even for comparatively small molecular weights.
A promising alternative for addressing the equilibrium issues is
based on the methods of statistical mechanics and, particularly, on the method
of integral equations of the theory of liquids commonly known as RISM
\cite{Chandler72,Stell,Monson90}.
The main advantage of the method based on the
Reference Interaction Site Model (RISM) equations 
is that here one explicitly deals with
all quantities averaged over the equilibrium Gibbs distribution.
This results in a considerable reduction of computational expenses for
many problems compared to those of direct simulation methods.
An example of application of the RISM technique to hydration of
biomolecules is the calculation of the alanine dipeptide carried out
in Ref. \onlinecite{Pettitt86}.

Practically, in the limit of infinite dilution the RISM method requires
numerical solution of a system of integral equations for the solute--solvent
pair correlation functions of order proportional to the number of
atoms in a macromolecule.
Here the solvent--solvent correlation functions can be found beforehand
and the conformation of a macromolecule is assumed fixed.
We note that memory requirements for keeping the matrix of unknown
variables could be quite considerable --- a few terrabytes of RAM or
more even for a DNA duplex of only 5 bases.

A number of coarse--grained models \cite{Hirata,Svensson}
for complicated macromolecules have
been suggested to eliminate this difficulty in the RISM method.
For instance, in Refs. \onlinecite{Hirata} the solvation of model DNA structures
has been investigated using such coarse--graining
with each nucleotide pair being approximated by a
potential centre and the solvent by a binary mixture of charged particles.
Similar coarse--grained models may be of interest for addressing non--specific
hydration effects, which are fairly independent of the detailed
molecular structure and thus universal for a given class of biomolecules.
In Ref. \onlinecite{TikhNonPolar} the hydration of a linear nonpolar chain in
aqueous solution has been considered to elucidate the general behaviour
of hydration of long macromolecular chains.
Unfortunately, errors resulting from a coarse--graining procedure are
practically impossible to estimate and therefore such models seem to be
too remote from the complexities of the real DNA molecule.

Here we suggest an efficient numerical algorithm for solving the RISM
equations that makes it feasible to study a detailed model for short
dupleces of DNA in water in a quite moderate computational time, even on
a PC and, most importantly, avoiding the necessity to deal explicitly
with huge matrices.
The main idea of this algorithm is to 
use the Newton--Raphson scheme with a specially appromixated Jacobi matrix
with consequent application of the
nonstationary iterative methods for solving large systems of
linear equations. A review of modern developments in the latter field,
on which our approach heavily relies, may be found in Ref. \onlinecite{KelleyBook}.

In this work we describe the new algorithm for calculation of the
solvent--solute correlation functions of a macromolecule in aqueous
solution. This is then applied to study of the hydration of
B--DNA dupleces.

\section{Basic equations of the RISM theory}
\label{sec:eqs}

Various integral equation techniques have proved to be successful for a wide
range of applications from simple monoatomic liquids to complex
multicomponent mixtures of molecular liquids.
Generally, the correlation functions satisfy the infinite chain of
Bogoliubov--Born--Green--Kirkwood--Yvon (BBGKY)
integro--differential equations.
To achieve a practical progress one possibility is to apply a closure relation
expressing high--order correlation functions in terms of a few low--order
ones. Many different closure relations have been suggested, each having
its own region of validity, which can be established by comparing the
results of calculations with available data from experiment or
computer simulations.

Another particularly fruitful strategy is based on the Ornstein--Zernike
(OZ) equation, which relates the direct, $\tilde{C}(k)$, and the total,
$\tilde{h}(k)$, correlation functions in the Fourier space,
\begin{equation}
(1\,-\,\rho\tilde{C}(k)\,)\,(1\,+\,\rho\tilde{h}(k)\,)\,=\,1,
\label{simply_oz}
\end{equation}
where $\rho$ is the density of the liquid.
Here the original and the Fourier conjugated functions, the latter being
designated by the tilde, are related as follows,
\begin{equation}\label{eq:fourier}
\tilde{x}(k) = \frac{4\pi}{k}\int_{0}^{\infty} rdr\,\sin(kr)\,x(r),\quad
x(r) = \int_{0}^{\infty} kdk\,\frac{\sin(kr)}{2\pi^2 r} \,\tilde{x}(k).
\end{equation}
The total correlation function is defined in such a way 
that $h(r) + 1$ is the
probability density to find an atom at the distance $r$ from another one.
Eq.~(\ref{simply_oz}) is an exact integral equation between $h(r)$
and the direct correlation function, $C(r)$, which can be established 
based on the diagrammatic cluster expansion. This has a simple physical meaning.
$C(R)$ is said to be caused by the two--body interaction
potential between a pair of atoms, $u(r)$, whereas the indirect correlation
function, $\gamma(r) = h(r) - C(r)$, contains contributions of all other
atoms.

The direct correlation function also has a
diagrammatic representation, which can be viewed as the exact closure relation
between the two unknown functions. However, 
one would like to express such a relation in an explicit form
which is popularly written as,
\begin{equation} \label{closureRel}
C(r)\,=\,\exp(-u(r)/k_{\rm B}T+\gamma(r)+{\cal B}[\gamma(r)])\,-\,
\gamma(r)\,-\,1,\label{funk}
\end{equation}
where $k_{\rm B}$ is the Boltzmann constant, $T$ is the absolute temperature
and ${\cal B}[\gamma(r)]$ is called the bridge functional.
Unfortunately, no strict analytical expression can be found for the latter.

Therefore,
in practice, one has to revert to some kind of an approximate closure relation
inspired by partial resummations of the diagrammatic expansion \cite{Monson90}.
Let us mention here two of the most widely used relations,
\begin{eqnarray}
{\cal B}[\gamma(r)] &=& 0\label{hnc},\\
{\cal B}[\gamma(r)] &=& \ln(1+\gamma(r))-\gamma(r)\label{py},
\end{eqnarray}
which are called the hyper--netted chain (HNC) and the Perkus--Yevick (PY)
closures respectively.
It is well established that, generally, the approximation (\ref{hnc}) is
more accurate for long--range potentials, whilst approximation (\ref{py})
works better for short--range potentials.

Generalisation of the OZ equation to multicomponent mixtures of molecular
liquids seems to be relatively straightforward. 
The conformation of a molecule is assumed to
be fixed, so that the correlation function between the $i$-th and $j$-th
sites (atoms) inside the same molecule are given by
$v_{ij}(r)=\delta(r-r_{ij})/4\pi{}r^2$, where $r_{ij}$ are
the known interatomic distances.
In the Fourier space the analogue of Eq.~(\ref{simply_oz})
can be written in the matrix form,
\begin{equation}
\left(\,\hat{v}(k)+\hat{\rho} \hat{h}(k)\,\right)\,
\left(\,\hat{v}(k)^{-1}-\hat{\rho} \hat{c}(k)\,\right)\,=\,\hat{1},
\label{simply_rism}
\end{equation}
where $\hat{v}$ is a block diagonal matrix with the number of blocks equal
to the number of distinct molecules and the elements of each block are
equal to $v_{ij}(k)=\sin(kr_{ij})/kr_{ij}$. 
The matrix $\hat{\rho}$ contains the partial atomic densities of
the mixture and possesses the same block structure as the matrix $\hat{v}$.
The form (\ref{simply_rism}) is often called the Site-Site Ornstein--Zernike
equation (SSOZ), and jointly with a particular closure relation 
(\ref{closureRel}) such as e.g. HNC or PY is loosely referred to as the RISM
theory. We shall favour this terminology for historical reasons
despite it may not be the most logical one \cite{Monson90}.

The important point is, however, that the relation of these equations
to the diagrammatic expansion is no longer preserved in the site--site
version of the formalism. So, in a sense, the SSOZ equation may 
be simply viewed as a formal definition of the site--site direct 
correlation function \cite{Stell}, with no particular deep meaning attached to
the analogues of HNC or PY closures.
The shortcomings of the RISM technique, that are possibly related to this
circumstance,
are well known and include the problems
with the long--ranged orientational correlations and the equation of
state (to discussion of these problems and attempts
to improve the RISM theory we refer the reader to standard reviews 
such as e.g. \onlinecite{Monson90}).
Despite these deficiencies, we believe that the RISM remains a very useful 
tool for studying hydration of complex biomolecules \cite{Hirata93}, where no 
other method can really compare to it in computational power.
However, based on the standard techniques 
in Ref. \cite{Hirata93} it was possible to consider the system of 
413 atoms only thanks to using a powerful supercomputer.

For a two--component molecular solution let us denote by $\rho$ the
density of the solvent and by $\rho_u$ the density of the solute.
In the limit of infinite dilution, $\rho_u/\rho \to 0$, Eq. (\ref{simply_rism})
is decoupled,
\begin{eqnarray}
\hat{\gamma}^v & = & \hat{v}^v \hat{c}^v \hat{v}^v\,
\left(\hat{1}-\rho \hat{c}^v\hat{v}^v\right)^{-1}\,-\,\hat{c}^v(k),
\label{rism_1}\\
\hat{\gamma}^{uv} & = & \hat{v}^u \hat{c}^{uv}
\left(\hat{v}^v+\rho \hat{h}^v\right)\,-\, \hat{c}^{uv},
\label{rism_2}
\end{eqnarray}
where the superscripts $v$ and $u$ refer to the solvent and the solute
respectively.

If we are interested in hydration of a $N$-atomic molecule, Eq. (\ref{rism_2})
for the solute--solvent correlation functions of all atoms in the molecule
with the O and H atoms of water may be rewritten in the Fourier
space as follows,
\begin{equation}
G\,=\,(\hat{V}\,\otimes\,\hat{W}-\hat{1}_{2N})\,C, \label{master}
\end{equation}
where $G$ and $C$ are vectors of size $2N$ containing the
correlation functions,
\begin{eqnarray}
G^T & = & \left[ \tilde{\gamma}_{1{\rm H}}(k), \ldots,
          \ \tilde{\gamma}_{N{\rm H}}(k),\ \tilde{\gamma}_{1{\rm O}}(k),
          \ldots,\ \tilde{\gamma}_{N{\rm O}}(k) \right], \label{ggg} \\
C^T & = & \left[ \tilde{c}_{1{\rm H}}(k), \ldots,
          \ \tilde{c}_{N{\rm H}}(k),\ \tilde{c}_{1{\rm O}}(k),
          \ldots,\ \tilde{c}_{N{\rm O}}(k) \right], \label{ccc}
\end{eqnarray}
where the superscript $T$ stands for the transposition.
Here the first subscript of a correlation function is the atom number in
the solute molecule and the second superscript designates the atom type
in the water molecule. The structure matrix, $\hat{W}$, of size $N$ is
expressed via the interatomic distances, $r_{ij}$,
\begin{equation}
W_{ij}=\frac{\sin(kr_{ij})}{kr_{ij}}.
\end{equation}
The matrix $\hat{V}$ expresses the geometrical and
correlational structure of the water,
\begin{equation}
\hat{V}=\left [\begin {array}{cc} v_{{1}}+v_{{2}}&
v_{{3}}\\\noalign{\medskip}2\,v_{{3}}&v_{{4}}\end {array}\right ],
\end{equation}
where its components are,
\begin{equation}
\begin{array}{ll}
v_{1} = 1 + \rho \tilde{h}_{\rm HH}(k), & 
v_{2} = {\sin(k\, r_{\rm HH})}/{k\, r_{\rm HH}} +
 \rho \tilde{h}_{\rm HH}(k), \\
v_{3} = {\sin(k\, r_{\rm OH})}/{k\, r_{\rm OH}} + \rho 
\tilde{h}_{\rm OH}(k),&
v_{4} = 1 + \rho \tilde{h}_{\rm OO}(k). 
\end{array}
\end{equation}
Here $r_{{\rm HH}}$ and $r_{{\rm OH}}$ are the interatomic distances in the
water molecule, $\rho$ is the density of water and
$\tilde{h}_{\rm HH}$, $\tilde{h}_{\rm OH}$ and $\tilde{h}_{\rm OO}$ are
the total solvent--solvent correlation functions which may be found
beforehand.

To avoid divergences we apply the HNC closure relation (\ref{hnc}) using
the renormalization procedure of the Coulomb potential proposed by Ng
\cite{Ng74,Monson90},
\begin{equation}
c^s_{i{\rm X}}(r)=\exp\left(-U_{i{\rm X}}(r)/k_{\rm B}T+f_{i{\rm X}}
(r)+\gamma^s_{i{\rm X}}(r)
\right)-\gamma^s_{i{\rm X}}(r)-1
\label{closure},
\end{equation}  
where $U_{i{\rm X}}(r)$ is the non--Coulombic part of the interatomic
potential, the subscript $i$ is the atom number in the solute molecule and
index X refers to atoms O or H in the water molecule. Here
the function $f_{i{\rm X}}$ is the renormalized Coulomb potential
expressed through the charges $q_i$ and $q_{\rm X}$: 
\begin{equation}
f_{i{\rm X}}(r) = -q_{i} q_{\rm X} \frac{1-\erf(r)}{rk_{\rm B}T}.
\end{equation}
The short--range parts of the correlation functions, which are designated by
the superscript $s$ above, are related to the complete correlation
functions as follows,
\begin{equation}
\tilde{\gamma}_{i{\rm X}}(k)=\tilde{\gamma}^s_{i{\rm X}}(k)-
\tilde{f}_{i{\rm X}}(k), \quad
\tilde{c}_{i{\rm X}}(k)=\tilde{c}^s_{i{\rm X}}(k)+\tilde{f}_{i{\rm X}}(k).
\end{equation}  
We also note that the introduction of the short--range correlation functions
allows one to perform the Fourier transformation using the fast Fourier
transform technique.

\section{The numerical algorithm}
\label{sec:num}

In this section we propose an algorithm for numerical solution of the system
of $2N$ integral equation (\ref{master}, \ref{closure}) for large values
of $N$.
We assume that the solvent density, $\rho$, the temperature $k_{\rm B}T$,
the matrices $\hat{W}$ and $\hat{V}$ and the interaction potentials
appearing in the closure relation (\ref{closure}) are given.

First of all, let us introduce discretisation by choosing $L$ grid points
uniformly distributed in the $r$- and $k$-spaces,
\begin{equation} \label{grid}
\Delta{r}\Delta{k} = \frac{\pi}{L}, \quad r_i = i\Delta{r},
\quad k_j = j\Delta{k}, \quad (i,j = 1\ldots L).
\end{equation}
We denote by $\vec{X}$ the $2NL$-component vector, which approximates
the set of functions, $\tilde{\gamma}_{iX}^{s}(k)$ in $j\Delta k$ grid points,
\begin{equation} \label{xxx}
X^T = \left[ \tilde{\gamma}^{s}_{1{\rm H}}(\Delta k),\ldots,
      \ \tilde{\gamma}^{s}_{1{\rm H}}(L\Delta k),\ldots,
      \ \tilde{\gamma}^{s}_{N{\rm O}}(L\Delta k) \right].
\end{equation}
Then we introduce the discrete Fourier transformation operators
${\cal F}^f$ and ${\cal F}^b$,
\begin{equation} \label{fffb}
\hat{F}^f = \frac{4\pi(\Delta r)^2}{\Delta k}\hat{S}, \quad
\hat{F}^b = \frac{(\Delta k)^2}{2\pi^2\Delta r}\hat{S}, \quad
S_{ij} = \frac{j}{i}\sin(\frac{\pi}{L}ij), \quad (i,j=1,\ldots,L).
\end{equation}
Eq. (\ref{master}) becomes a system of $2NL$ nonlinear algebraic equations,
\begin{equation}
\vec{X} - \vec{Z}[\vec{X}] \equiv \vec{X} - \hat{A}\,(\hat{I}_{2N}
\otimes \hat{F}^f) \vec{C}[(\hat{1}_{2N}\otimes \hat{F}^b)
\vec{X} ] -\vec{Q} = 0.
\label{dmaster}
\end{equation}
Here $\hat{A}$ is a matrix of size $2NL$ consisting of diagonal
blocks of size $L$, which is the discrete analogue of the matrix
$\hat{V}\otimes \hat{W}-\hat{1}_{2N}$, and
$\vec{Q} = (\hat{V}\,\otimes\,\hat{W})
[ \tilde{f}_{1{\rm H}}(k),\ldots,\ \tilde{f}_{N{\rm O}}(k)]^T$.
For discretisation of the closure relation (\ref{closure})
let us also introduce the function $\vec{C}[\,\vec{Y}]$,
\begin{equation} \label{dclosure}
C_{l} \equiv C_{l}[Y_{l}] = M_{l}\exp(Y_l)-Y_l-1,
 \quad (l=1,\ldots,2NL),
\end{equation}
where $M_l$ are elements of a $2NL$-dimensional vector $\vec{M}$,
\begin{equation} \label{mmml}
M^T = \left[ \exp(f_{1{\rm H}}(r)-U_{1{\rm H}}(r)), \ldots,
      \ \exp(f_{N{\rm O}}(r)-U_{N{\rm O}}(r)) \right].
\end{equation}

The simplest algorithm for solving the system of nonlinear algebraic
equations (\ref{dmaster}) would be the Picard method of direct iterations.
Thus, given an initial approximation, $\vec{X}^{(0)}$, the improved solution 
is found from the recurrent relation,
\begin{equation} \label{direct}
\vec{X}^{(n+1)}\,=\,\vec{Z}[\,\vec{X}^{(n)}\,], \quad (n=0,1,\ldots).
\end{equation}
The iterations should stop when the norm of $\vec{X}^{(n+1)}-\vec{X}^{(n)}$
becomes smaller than a predefined small value.
Some modifications of the method of direct iterations are often utilized,
which involve several previous iterations for providing a better
convergence. A good example of such modifications, based on a vector
extrapolation, was used in Ref. \onlinecite{Homeier95}.
The direct iteration method for solving of the RISM equations was used in
Ref. \onlinecite{Talitskih95}, where the principle of ``monotonic discrepancy''
was applied to improve convergence.
Unfortunately, in practice the convergence of the direct iteration method
seems to be an exception rather than the rule.

The Newton--Raphson algorithm appears to be much more efficient for
solving equations (\ref{dmaster}). Here iterations are performed by
the following scheme,
\begin{eqnarray}
\vec{X}^{(n+1)} & = & \vec{X}^{(n)}\,+\,\Delta\vec{X}^{(n)},\\
\label{newton}
\Delta\vec{X}^{(n)} & = & \hat{J}^{-1}\,
(\vec{Z}[\vec{X}^{n}] - \vec{X}^{n}),
\label{linsys}
\end{eqnarray}
where $\hat{J}$ is the Jacobi matrix obtained by differentiating
Eq. (\ref{dmaster}):
\begin{equation}
\hat{J}\,\equiv\,\hat{1}_{2NL}\,-\,\frac{d\vec{Z}}{d\vec{X}}\,\equiv\,
\hat{1}_{2NL}\,-\,\frac{2}{L}\,\hat{A}\,(\hat{1}_{2N}\otimes\hat{S})\,\hat{D}\,
(\hat{1}_{2N}\otimes\hat{S}).
\label{jacobi}
\end{equation}
The elements of the diagonal matrix $\hat{D}$ are given by,
\begin{equation}
D_{ll} = M_{l}\exp(Y_l)-1,\quad (l=1,\ldots,\,2NL).
\end{equation}
The disadvantage of the Newton--Raphson method is that at each iteration
step it is necessary to solve a system of linear equations (\ref{linsys})
of size $2NL$.
It is a rather large system, since, even in the case of comparatively small
$N$, the number of grid points $L$ should be of order of thousands to achieve a
good discretization.
Despite that the convergence of the Newton--Raphson iterations is rather
fast, in practice the expenses for solving the system of linear equations
with the Jacobi matrix of size $2NL$ are extremely high.

Gillan \cite{Gillan} has suggested a more efficient algorithm which
is essentially a hybrid of the Newton--Raphson and the Picard schemes for the
`coarse' and `fine' parts of the solution respectively, 
the former being obtained
as an expansion in the basis of so--called roof functions. 
In Ref. \onlinecite{Labik} it was noted that the expansion of the `coarse'
part as a truncated Fourier series yields even a better convergence.
Thus, the main feature of both the Gillan and the LMV methods is that
the solution is sought in a cycle of the Newton--Raphson steps for a
small--size system of equations for the expansion coefficients of 
the `coarse' part,
and a consequent Picard refinement for the solution. 
We note, however, that this attractive scheme \cite{Haymet}
is difficult to apply in practice for large molecules. 
Indeed, if the size of the system of equations for the
`coarse' part is taken relatively small, the consequent direct Picard iteration
would not converge. If, on the contrary, this size is taken sufficiently large,
the corresponding Jacobi matrix would not fit into the computer memory.

Here propose a somewhat different strategy which could be viewed as the
Newton--Raphson scheme with an appromixated Jacobi matrix.
Namely, let us choose a subset of the grid points in the coordinate
and $k$- spaces,
\begin{equation}
r^{'}_i = i\,m\,\Delta{r},\quad k^{'}_j = j\Delta{k}, \qquad 
(i,j = 1\ldots \frac{L}{m}), \label{grid1}
\end{equation}
where $m$ is an integer number.
The Jacobi matrix calculated on this subset is denoted as $\hat{J}_m$
and its size is $2NL/m$.
If we denote as $k^{''}$ the complementary subset of components,
we suggest to perform iterations according to the following formula,
\begin{eqnarray}
\vec{R}^{(n)}\,&=&\,\vec{X}^{(n)}\,-\,\vec{Z}[\vec{X}^{(n)}],
\label{resudial}\\
\vec{X}^{(n+1)}(k^{'})\,&=&\,\vec{X}^{(n)}(k^{'})\,-\,\hat{J}^{-1}_{m}\,
\vec{R}^{(n)}(k^{'}),
\label{newt}\\
\vec{X}^{(n+1)}(k^{''})\,&=&\,\vec{X}^{(n)}(k^{''})\,-\,\vec{R}^{(n)}(k^{''}).
\label{pic}
\end{eqnarray}
Thus, the approximated inverse Jacobi matrix has the elements of
the exact inverse to the reduced Jacobi matrix for the $k'$ subset indices,
and the unity matrix for the $k''$ subset indices.
Again, iterations are carried out until the norm of the vector $\vec{R}^{(n)}$
becomes smaller than a predefined small value.

The main feature of this particular scheme, which is an alternative to
the Gillan and LMW methods, is that, in a sense, it combines the
Newton--Rapson and Picard iterative schemes performed simultaneously.
The reduction parameter $m$, which could be chosen as an integer power of 2,
can be decreased during iterations if the convergence is slowing down,
or can be increased if the convergence is improving.
Practically, to achieve a good initial convergence, the size of the 
reduced Jacobi matrix $2NL/m$ is still going to be rather large for
the molecule we are going to study here.
Fortunately, now this difficulty can be eliminated.

For calculation of the inverse matrix in Eq. (\ref{newt}),
i.e. solving a system of linear equations, we propose to apply
the nonstationary iterative methods.
Basically, these methods are versions of the
conjugated gradients method. The main advantage of these methods is a better
convergence for a wider class of matrices.
Also, compared to the Gauss method, they do not require to keep the
matrix of the system in the computer memory. 
Instead, it would suffice here to
provide a procedure of matrix--vector multiplication,
$\vec{y}=\hat{J}\vec{x}$, for arbitrary vector $\vec{x}$.
If the number of grid points $L$ is equal to an integer power of 2,
then one can use the fast Fourier transform technique for calculation of
vector--matrix multiplications. It is straightforward to see that this
procedure requires $LN^2$ floating point operations for large $N$.
The additional calculational efforts per iteration in the nonstationary
iteration methods are proportional to $NL$. For example, the $Bi-CGSTAB$
method \cite{KelleyBook} requires two calculations of vector--matrix products,
and an additional $48NL$ floating point operations. 
Overall, the memory requirements
here are restricted to $20NL$ floating point units.

\section{Numerical results}
\label{sec:res}

In this section
we study fragments of the double helix DNA formed by nucleotide pairs
G:C. We have calculated fragments of different lengths (from 1 to 5 pairs),
so that the number of atoms $N$ varied from 63 to 315.

We assumed the geometry of these fragments to be standard \cite{Landoldt89}
and the interaction potentials of atoms in different pairs in a fragment
with water equal.
According to Ref. \onlinecite{Poltev92} the non--Coulombic part of the interaction
potentials of DNA atoms with water atoms can be taken in the form,
\begin{equation}
U_{iX}(r)\,=\,A_{iX}/r^{12}\,-\,B_{iX}/r^6.
\label{pot1}
\end{equation}
If an atom forms a hydrogen bond we use another expression
for the potential,
\begin{equation}
U_{iX}(r)\,=\,A'_{iX}/r^{12}\,-\,B'_{iX}/r^{10}.
\label{pot2}
\end{equation}
The latter form is used in the AMBER force field potential parametrisation
\cite{AMBER}. It is added to merely fine--tune the distances between atoms 
forming hydrogen bonds, which, in principle, are well enough represented by the
Coulombic attraction alone.
The potential parameters were taken from Ref. \onlinecite{Poltev92}
and the atomic charges were taken from Ref. \onlinecite{Jurkin80}.
As solvent we use water at the normal density $1\ g/sm^3$ and
the temperature $25^{o}C$.
Correlation functions of pure water were calculated using Eq.~(\ref{rism_1})
in the HNC closure for the potential model TIPS with the parameters
taken from Ref.~\onlinecite{Jorgensen81}.
The charge of the oxygen atom in the water molecule we consider to be
equal to $q_{\rm O} = -0.86$ and that of the hydrogen atom --- to
$q_{\rm H}=-q_{\rm O}/2$.
The interatomic distances in the DNA fragment were approximated up to
the precision of $0.01\,A$ to reduce the number of distinct elements
in the matrix $\hat{W}$. This allowed us to strongly reduce the
requirements for RAM. Nevertheless, our calculations show that the
correlation functions change rather insignificantly due to this round--off.
The number of grid points and step size were taken equal to $L = 1024$ and
$\Delta r=0.025\,A$. The reduction parameter was set to $m = 8$, so that
the size of the linear system is equal to 80640.
The $TFQMR$ nonstationary method \cite{KelleyBook} was applied
for solving the reduced system of linear equations.
The precision parameters for solution of the nonlinear equations,
$\epsilon$, and of the linear ones, $\delta$, were equal to $10^{-5}$
and $10^{-6}$ respectively.
Calculations were carried out as follows. First, we calculate the
correlation functions of one nucleotide pair. The output of this
calculation is then used as input for calculation of two nucleotide
pairs, and so on.

In Tab.~\ref{tab:exp} we present the computational expense of the problem.
One can see that the computational time does not grow
catastrophically with the increasing system size.
The number of iterations, $N^{ITER}$, remains sufficiently low for
fragments lengths considered here. This is because
the calculational time is determined mainly by the time required for
matrix--vector multiplications and it grows approximately as the second
power of the reduced linear system size, $2NL/m$.

In Fig.~1 we exhibit the calculated correlation functions, $h(r) + 1$,
for some of the DNA duplex atoms with atoms of water.
Heights of the first maximum of the correlation functions vary from 4 to 0.4.
High peaks of correlation functions are observed for heteroatoms
in bases and for the oxygen atoms in the sugar--phosphate backbone.
Heights of the first peak, which are less than unity, correspond to those
atoms which are buried in the DNA molecule and thus screened by other atoms
from a direct contact with water molecules.

Now, let us consider separately the hydration of bases, the sugar--phosphate
backbone and also the characteristic features of hydration of the major
and minor grooves of DNA.

\subsection{Hydration of bases}

The highest correlation peaks correspond to the nitrogen atoms of guanine
N2 (see Fig.~1a) and N7 (Fig.~1b).
The position of the peak in the correlation function between
the hydrogen of water molecules and N7 in guanine is shifted
to the left compared to the peak of the correlation between the oxygen
of water and N7. 
The positions of the peaks in correlations atom N2 --- oxygen in water
and atom N2 --- hydrogen in water are nearly identical.
Also, the height of the first peak in correlations of
atom N2 --- hydrogen is much smaller than that for the atom N7.
According to the results from Monte Carlo simulations \cite{Teplukhin92}
this difference in behavior and degree of hydration of N7 and N2 is a
consequence of that N7 atom is a proton acceptor, while N2 is a proton
donor in the hydrogen bond. Obviously this results in different water molecule
orientations around these atoms, and this is manifested in the considered
correlation functions.
The characteristic difference in the hydration of these atoms along the
duplex chain is that the hydration of N7 does not change along the chain,
whilst the hydration of N2 increases from ends to the centre
approximately in 1.5 times.

The oxygen O2 in cytosine (see Fig.~1c) participates in a hydrogen bond
with an amino group of guanine through the atom 1H2.
The oxygen O6 in guanine (Fig.~1d) forms a hydrogen
bond with an amino group of cytosine through atom 2H4.
The correlation functions of these two atoms are similar to each other.
The characteristic feature of these functions is increasing of the first
peak heights from ends to the centre analogously to N2 atom in guanine.
Similarly, the oxygen atoms participate in forming hydrogen bonds with
water as can be seen from shifted peaks in their correlation functions
with the hydrogen of water molecules.

The hydrogen atoms in amino groups of guanine, 2H2 (Fig.~1e) and
1H2 (Fig.~1f), and 1H4 (Fig.~1g) and 2H4 (Fig.~1h) of cytosine
differ significantly in their hydration behaviour.
Protons participating in hydrogen bonds with oxygens O2 and O6 produce
approximately twice lower height of the first correlation peak compared
to that of the correlation functions for 2H2 of guanine and 1H4 of cytosine.
This is not surprising since hydrogens 2H2 and 1H4 are open for water
molecules, while 2H4 and 1H2 atoms hydrate weakly.

Hydration mechanisms of N1 atom in guanine (see Fig.~1i) and
N3 (Fig.~1j) atom in cytosine, which
both participate in hydrogen bonds between guanine and cytosine, are very
different from each other. The distinction can be observed in the height
of the first correlation peak with the hydrogen atom.
N3 atom in cytosine possesses a large partial charge (approximately $-0.7$),
resulting in a shift of the correlation peak with the hydrogen and in
its comparatively large height.
At the same time, the first peak in the correlation function with the oxygen is
somewhat lower than unity for inner nucleotides of the DNA duplex. This
means that, as in the case of the N1 atom, water weakly penetrates into the
inter--planar space of neighbouring nucleotide pairs.
 
N3 atom of guanine (Fig.~1k) is partially screened from water molecules by
amino group atoms. Thus, it hydrates in a much smaller degree than
N7 atom (Fig.~1b), though atom N3 possesses a larger partial charge ($-0.63$)
compared to the N7 atom ($-0.54$).
N9 nitrogen atom in guanine (Fig.~1l) and N1 atom in cytosine (Fig.~1m)
do not hydrate at all since they form a glycosidic bond.

\subsection{Hydration of the sugar--phosphate backbone}

Correlation functions of O1P atom of C residue on the
5' end are smaller than that in the other four nucleotide pairs (Fig.~1n).
On the contrary, the O2P atom of C-nucleotide on 5' end possesses
higher correlation peak compared to other pairs (Fig.~1o).
We can observe the same picture for oxygen atoms in the phosphate group
of G-nucleotide residue (Fig.~1p and Fig.~1q).

The phosphorus atoms in C and G residues hydrate nearly
equally, except for the 5' end atoms, where the first correlation peak
becomes broader (Fig.~1r and Fig.~1s).
As for the hydration of deoxyribose, we should note that the ring atom
O4* hydrates most strongly, the correlation peak on the 5' end being
smaller for the G residue (Fig.~1t and Fig.~1u).
O3* oxygens in G and C residues do not hydrate except
for atoms on 3' ends, which can be seen from Figs.~1v, 1w.
Hydration of G O5* (Fig.~1x) and C O5* (Fig.~1y) atoms does not
differ from each other. Unlike the O3* atom (Fig.~1v, 1w),
the hydration of end-- and center-- O5* atoms is the same.
The rest of carbon and hydrogen atoms in the sugar--phosphate backbone
hydrate rather weakly.

\subsection{Hydration of the major and the minor grooves}

Strong hydration observed for N7 (Fig.~1b) and O6 (Fig.~1d) atoms of
guanine and N4 atom of cytosine (Fig.~1z) qualitatively corresponds
to the hydration regions W1 and W2 presented in
Ref. \onlinecite{Schneider93}, while hydration of O2 atom in cytosine
(Fig.~1c) and N3 (Fig.~1k), N2 (Fig.~1a), 2H2 (Fig.~1e) atoms in
guanine correspond to the region S1.

On the other hand, as we have already mentioned, the first peaks in
the correlation functions grow from ends of the duplex to its center
for heteroatoms N2 (Fig.~1a), O6 (Fig.~1d), N4 (Fig.~1z) and O2 (Fig.~1c)
participating in hydrogen bonds between guanine and cytosine. 
On the other hand, it is known from the calculations of hydration for
monoatomic nonpolar chains \cite{TikhNonPolar} and can be explained from
the steric considerations, that the first peak of correlation function is
much higher for end--atoms than for center--atoms.
Such distinction in hydration behavior may be explained by two factors:
first, by the presence of large partial charges on heteroatoms, and also
by the geometrical structure of a Watson--Crick pair G:C.
Presumably, this effect has a cooperative origin and is connected to
the orientational ordering of water molecules in the grooves.

The values of correlation functions calculated by us are in good
agreement with the results of Ref. \onlinecite{Hummer94},
where the microscopic densities of oxygen and hydrogen atoms of water were
calculated in the region contiguous to the major and the minor grooves of
DNA using the mean--force potential formalism.

\section{Conclusion and discussions}
\label{sec:con}

In the current work we have developed an efficient algorithm for solving
the integral equations of the theory of liquids in the RISM approximation
for studying hydration of macromolecules.
The current treatment allows one to study hydration without applying
any additional simplifying assumptions about the molecular structure of
a macromolecule. The algorithm can be used to consider comparatively long
(several dozens of nucleotide pairs) biologically functional DNA fragments
using small computers.
It can also be used for studying hydration of promoters and
genome terminators.

In particular,
here we have studied the hydration of B-DNA fragments consisting of
identical nucleotide pairs.
Our results lead us to the conclusion that the hydration of nucleotide atoms
in the DNA chain depends mainly only on the structure of its neighbouring
environment, i.e. it is local by its nature.
Our conclusions may as well be extended
to an arbitrary sequence of G:C pairs. One expects that the hydration
of the sugar--phosphate backbone depends very weakly on the
particular sequence of base pairs. However, study of the dependence of the
hydration of nucleotide pairs in the chain on the type of neighbouring
pairs is beyond the scope of current paper.

In our calculations we assumed a rigid geometrical structure of DNA duplex.
There are reasons to believe that the hydration of a flexible or a bent
double helix differs from that of a rod--like structure.
We believe that the flexibility
of five nucleotide pairs is small enough for these effects to appear.
In principle, the RISM method allows one to calculate the mean--force
potentials as function of a particular molecular geometry, which would
finally permit study of the problem of the optimal DNA geometry in water
by Monte Carlo method \cite{Talitskih95}.

The reason we chose the length of the DNA duplex equal to five nucleotide
pairs is that, as our calculations show, for this length the influence of
the ends on the hydration of the central pair is very small.
This can be clearly seen from the data, since in most cases the correlation
functions of the penultimate and the central pairs are nearly identical.
Thus, one can assume that the hydration of the central nucleotide pair
in a short chain differs insignificantly from that of a nucleotide pair
inside an infinitely long chain. We note that in this respect the solvation
of monoatomic nonpolar chains in the aqueous environment is analogous to
that of DNA \cite{TikhNonPolar}.

We realize that the current molecular model is limited in many respects
since the DNA molecule was electro--neutral here due to a parametric
account of the phosphate screening.
Nevertheless, the existing experience of calculations of DNA hydration
by various methods indicates that this approximation appears to be quite
adequate in the region restricted by the first coordination sphere
\cite{Chuprina91,Teplukhin92,Hummer95}.
The method used by us can be extended to include free mobile ions
in the model and thus to study macromolecules in ionic solutions.

\acknowledgments

The authors thank Professor V.I.~Poltev and Dr O.A.~Mornev for
interesting discussions and useful suggestions and
Dr S.S.~Kolesnikov and Dr Yu.V.~Bobkov for technical assistance. 
We also acknowledge the support of the Centre for High Performance
Computing Applications, University College Dublin, Ireland.
This work was supported in part by grant INTAS-93-2084.


\begin{table}
\caption{Computational expense for calculation of the DNA hydration.
Here $N^{G:C}$ is the number of nucleotide pairs, $N$ is the number of
atoms in the fragment, $N^{W}$ is the number of distinct elements in the
structure matrix $\hat{W}$, $N^{TFQMR}$ is the number of iterations
required for solving the linear system, $N^{ITER}$ is the total number
of iterations for the given DNA fragment and $\tau_{PPro}$ is the
computational time in minutes for a Pentium PRO 200 MHz and 64 Mb
RAM computer.
}
$\left.\right.$\\
\label{tab:exp}
\begin{center}
\begin{tabular}{|r|r|r|r|r|r|}
\hline
$N^{G:C}$ & $N$  & $N^{W}$ & $N^{TFQMR}$ & $N^{ITER}$ & $\tau_{PPro}$ \\
\hline
 1       &  63  &   1038  &   124       &   7        &      12       \\
 2       & 126  &   1574  &   142       &   8        &      35       \\
 3       & 189  &   1715  &   149       &   8        &      70       \\
 4       & 252  &   1798  &   140       &   7        &     120       \\
 5       & 315  &   2020  &   148       &   7        &     190       \\
\hline
\end{tabular}
\end{center}
\end{table}

\begin{figure}
\caption{
Plots of the correlation functions, $h(r) + 1$, of some atoms in the
DNA duplex with atoms of water versus the radius, $r$,
measured in Angst\"oms. In each figure the solid line corresponds to the
correlation function of the atom with the hydrogen atom of water molecules
and the dashed line with points corresponds to the correlation function
of the atom with the oxygen atom of water molecules.
In each row we present from left to right the correlation functions
corresponding to the given atom in the first, second, third and fifth
nucleotide pairs of the B--DNA duplex d(GGGGG)$\cdot$d(CCCCC) starting from the
5' atom in guanine.
The data for the forth pair is suppressed.
}
\end{figure}

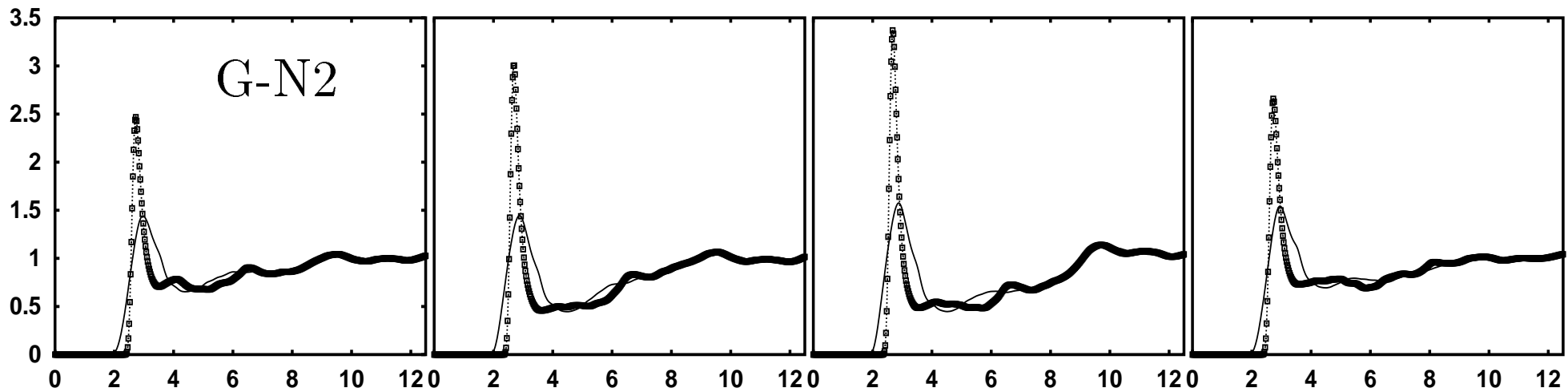
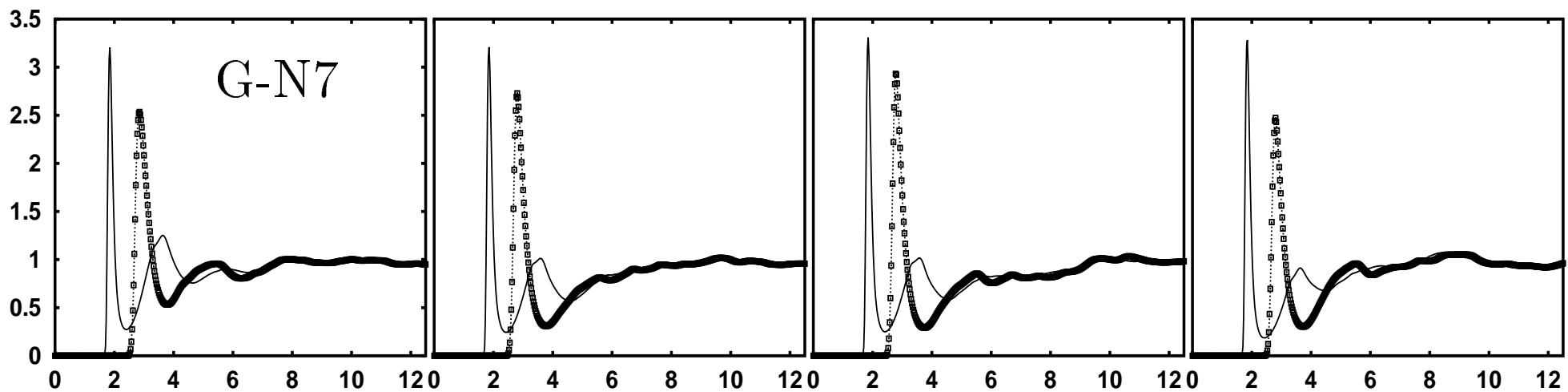
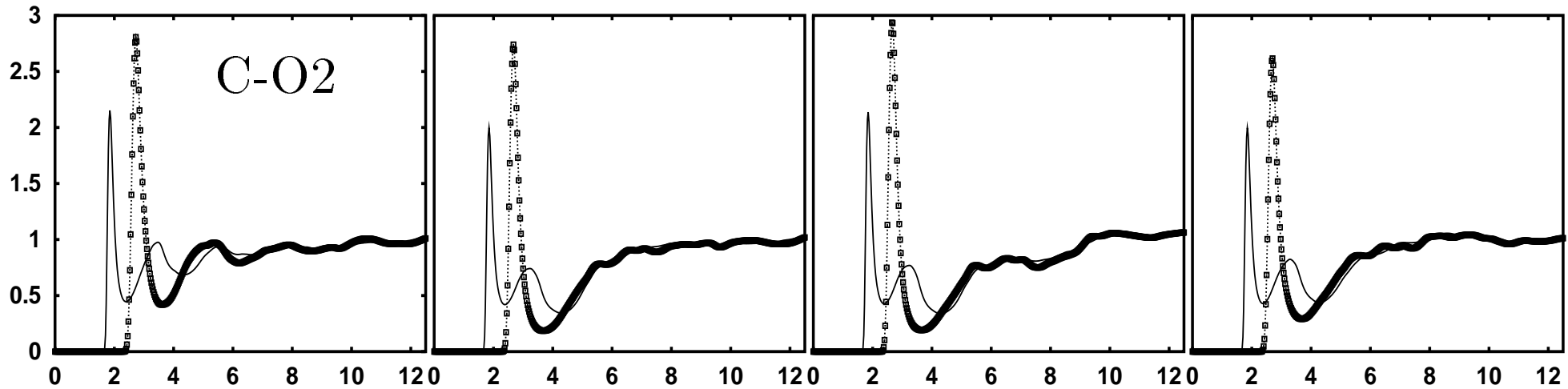
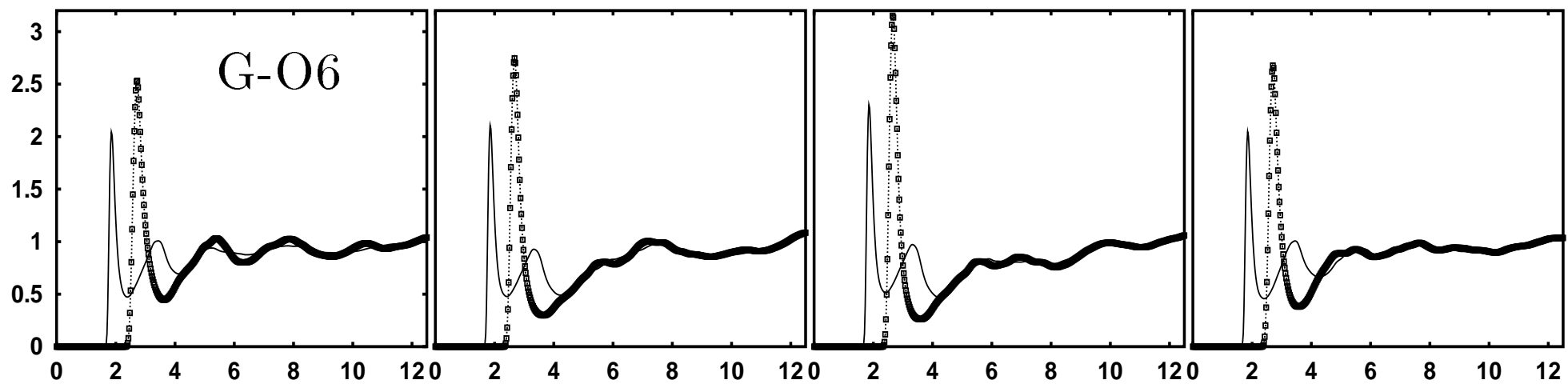
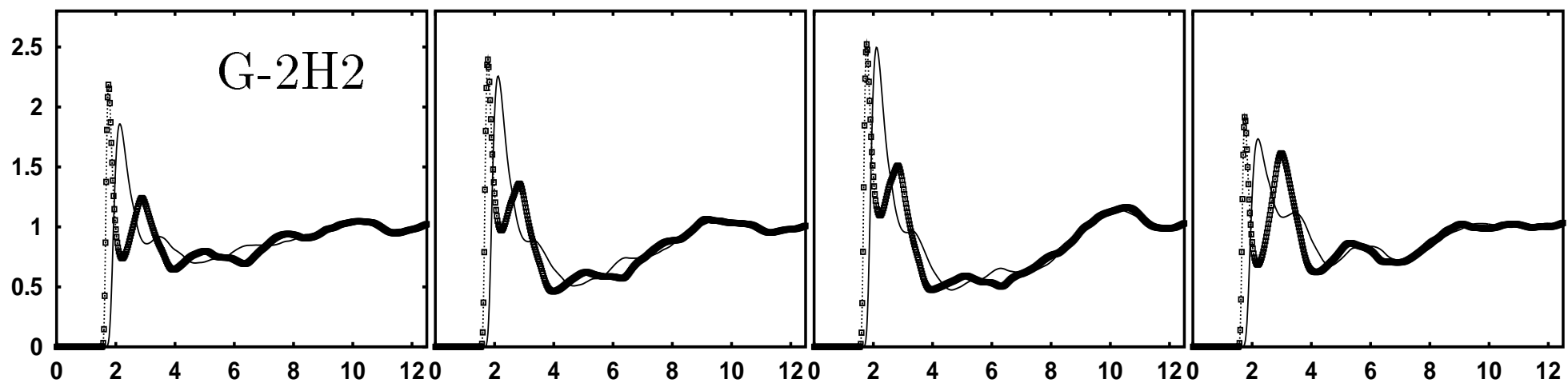
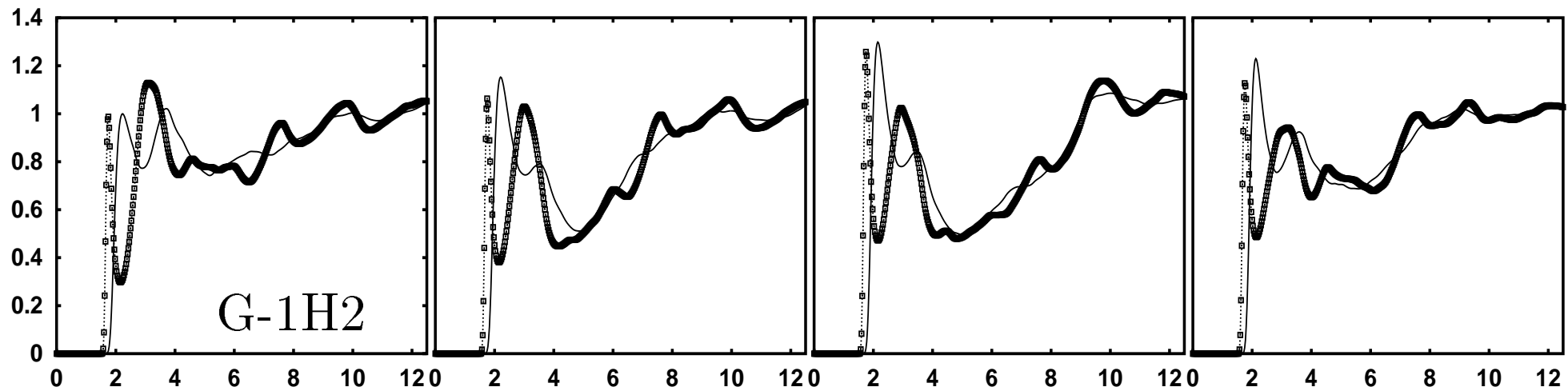
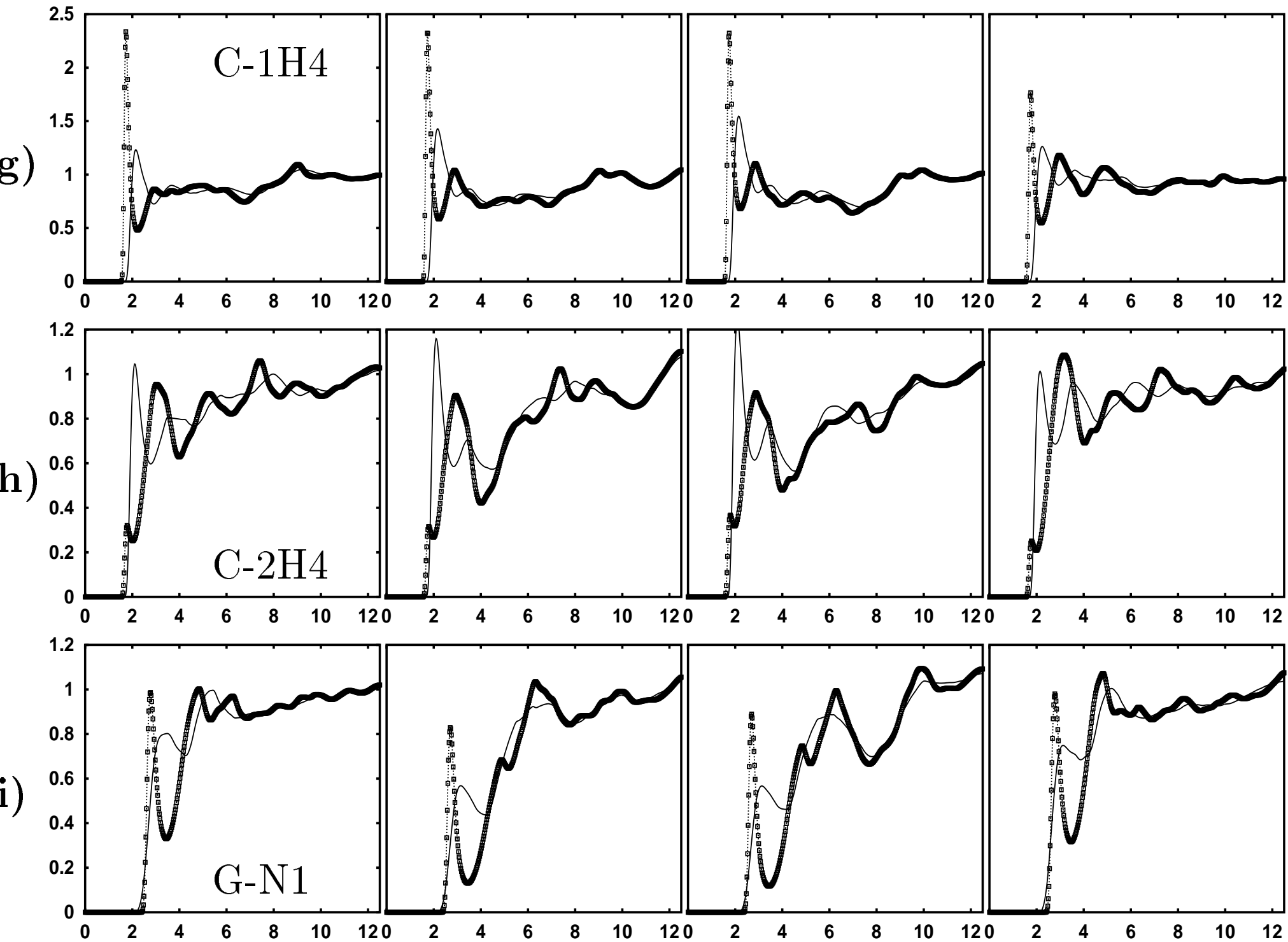
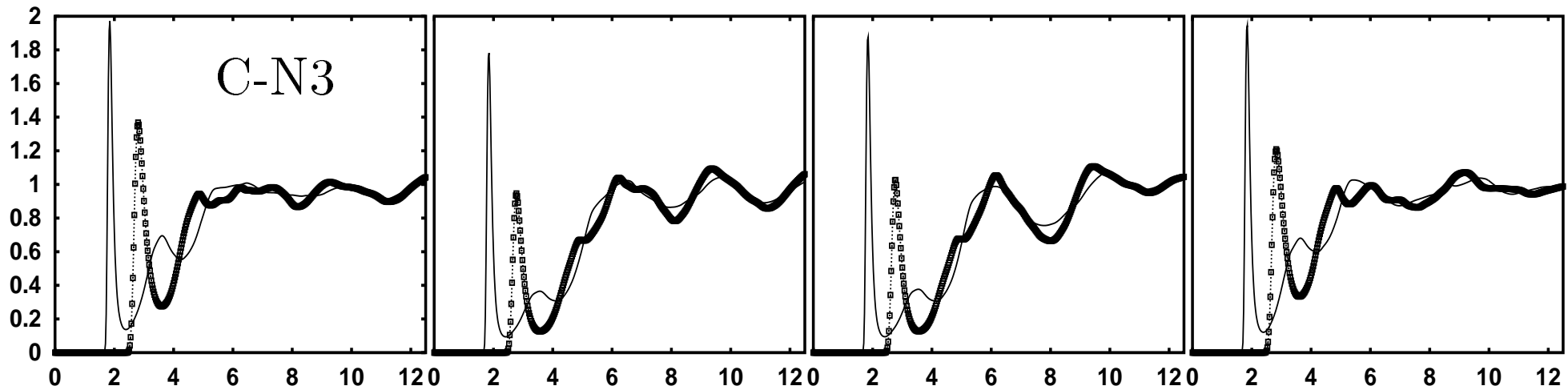
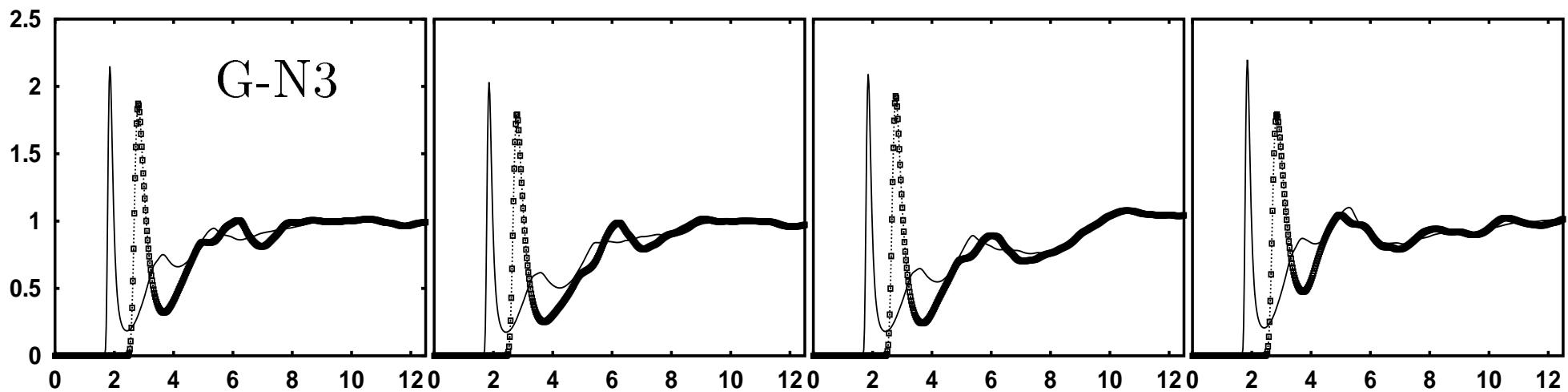
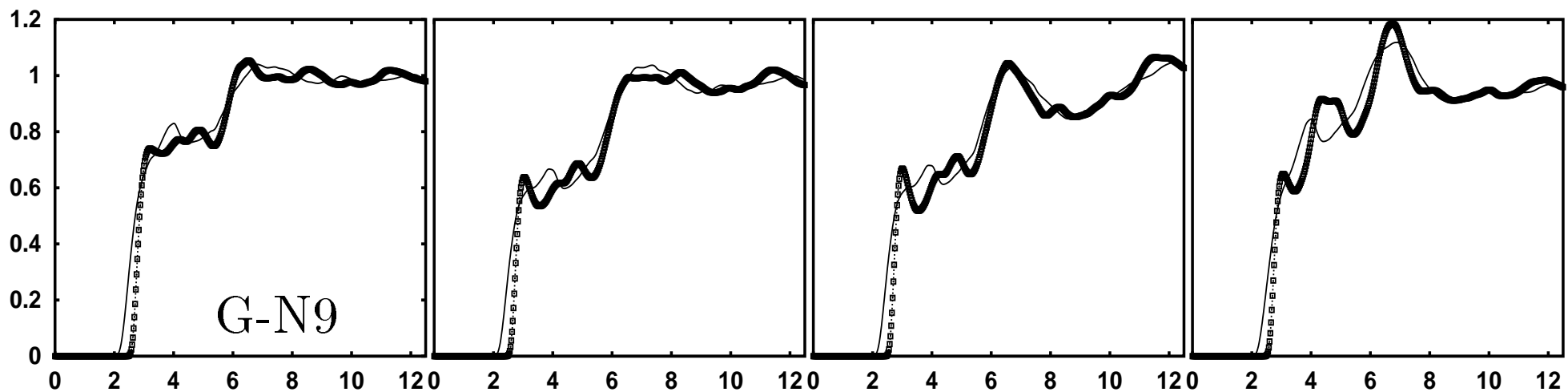
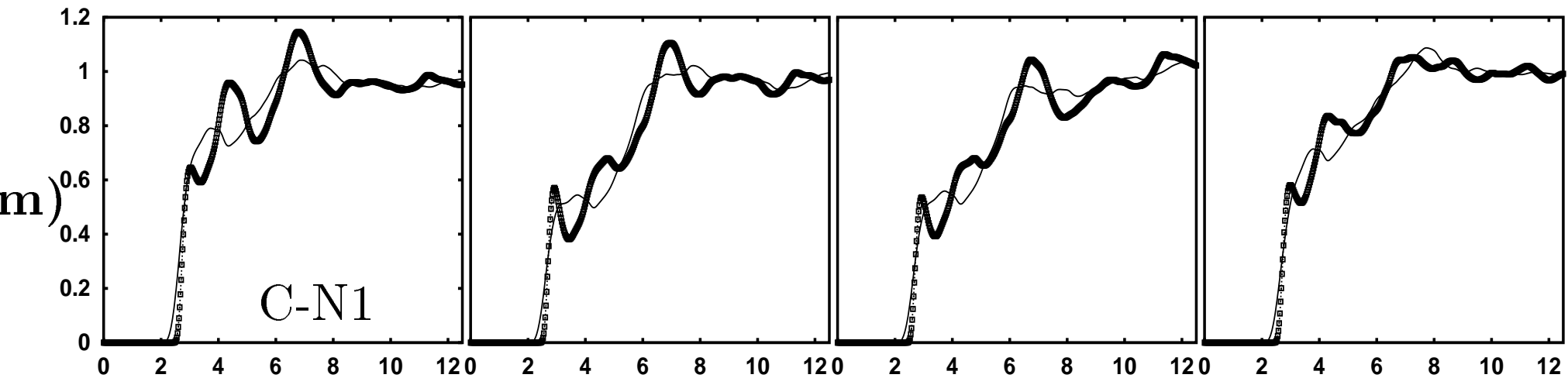
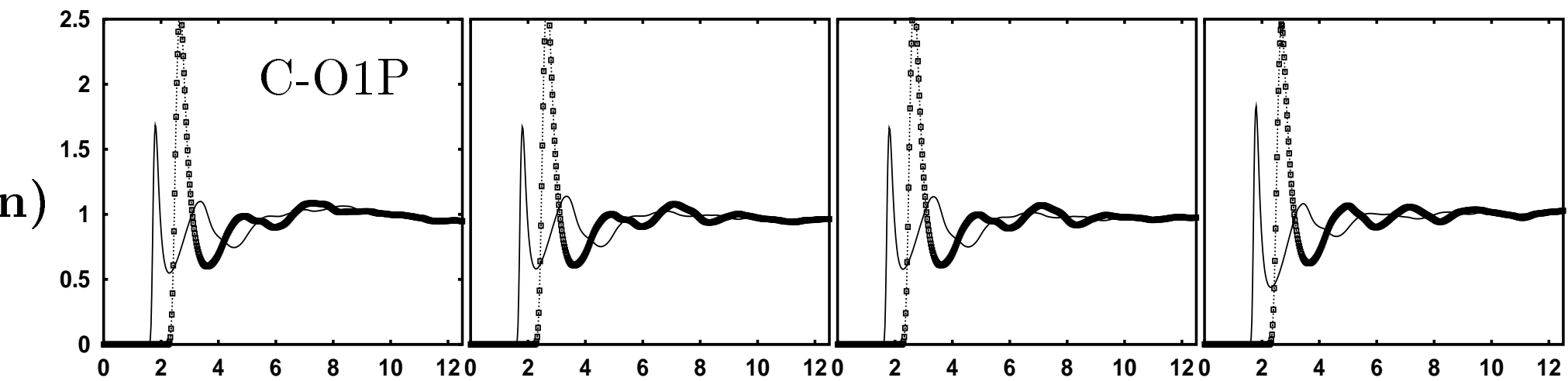
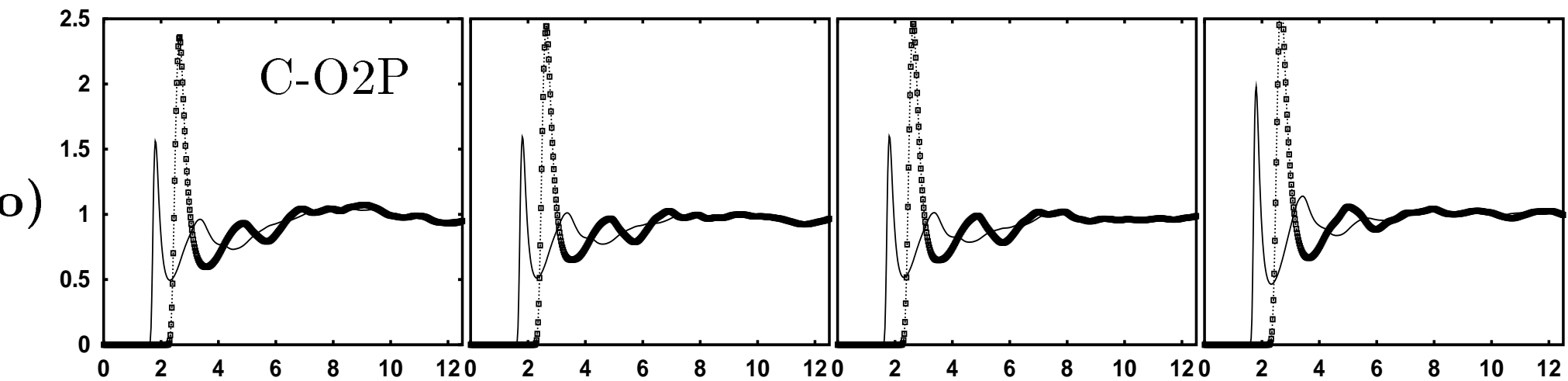
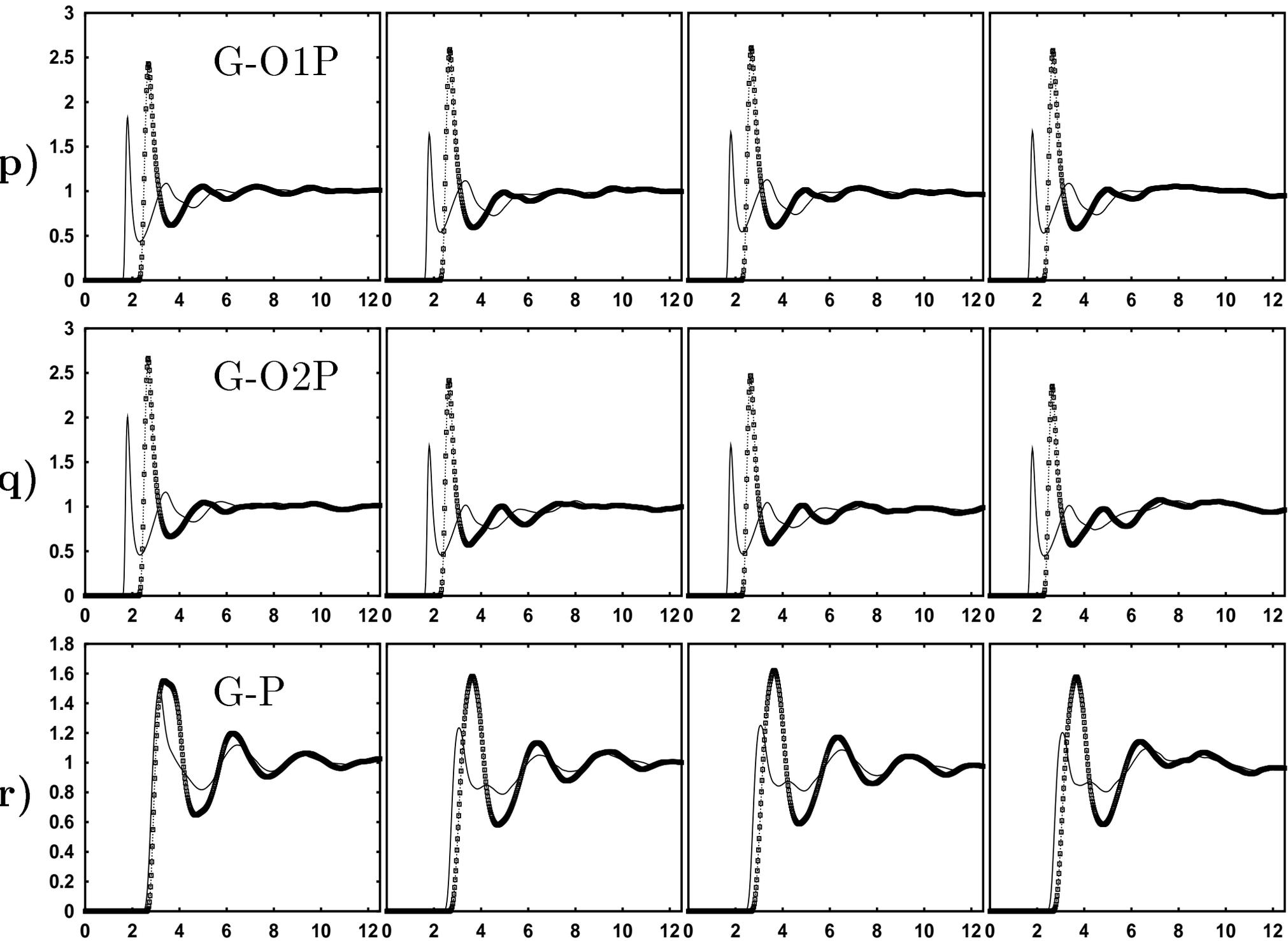
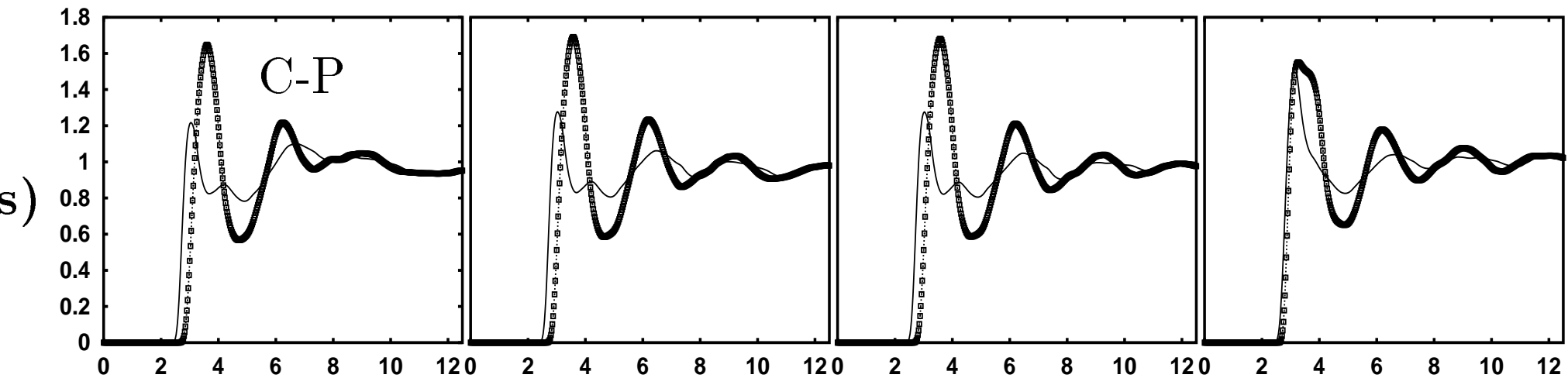
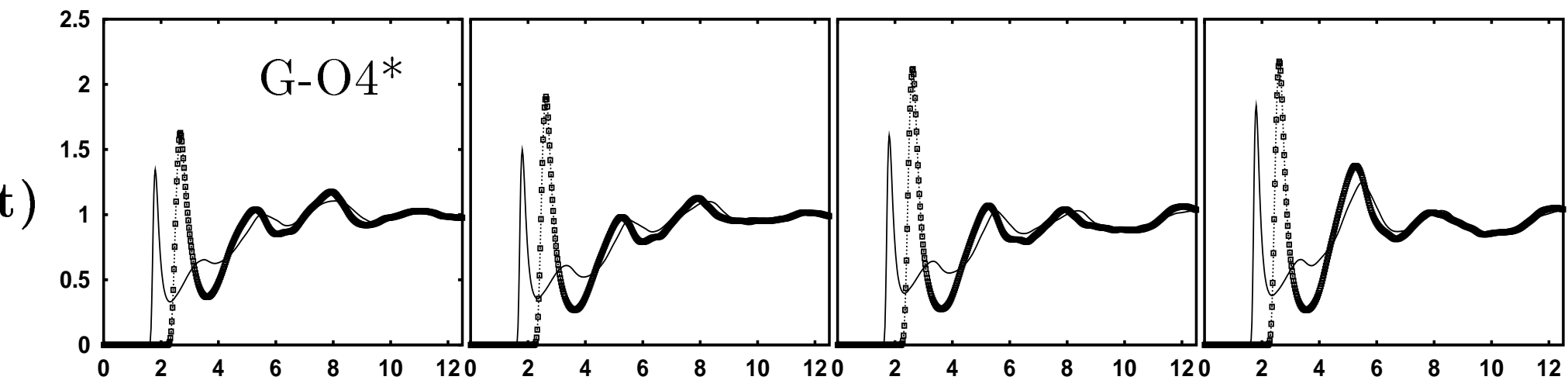
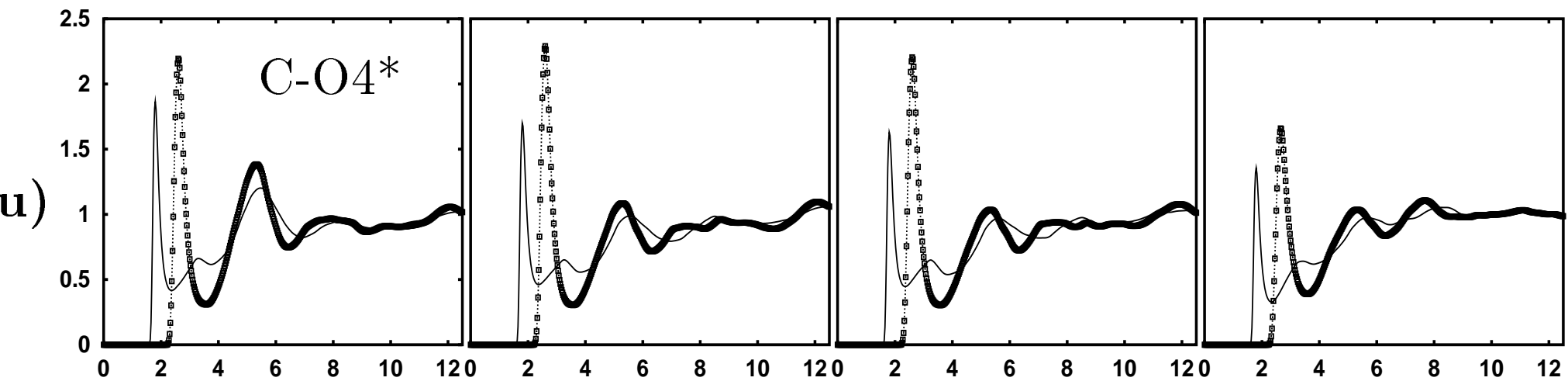
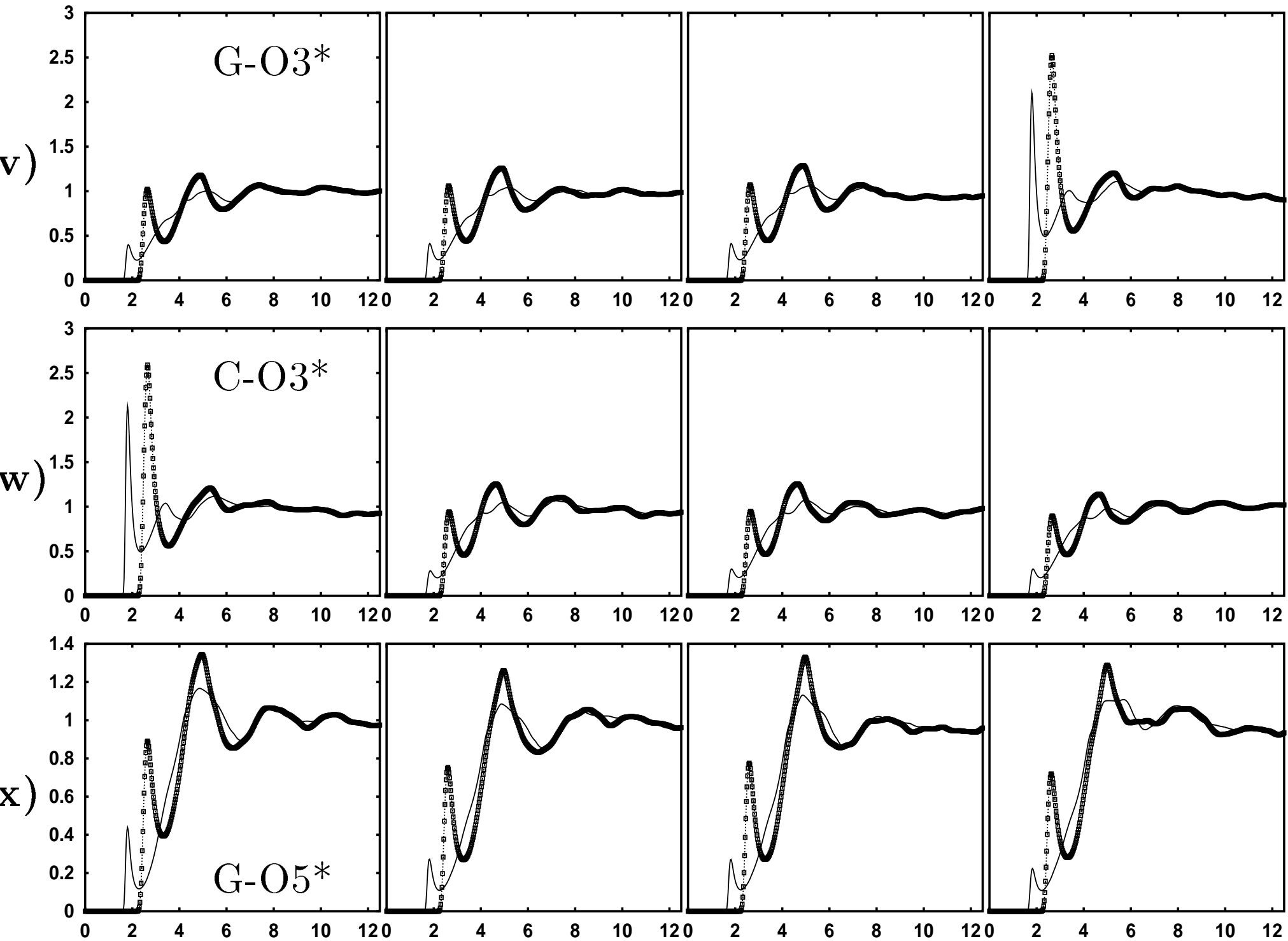
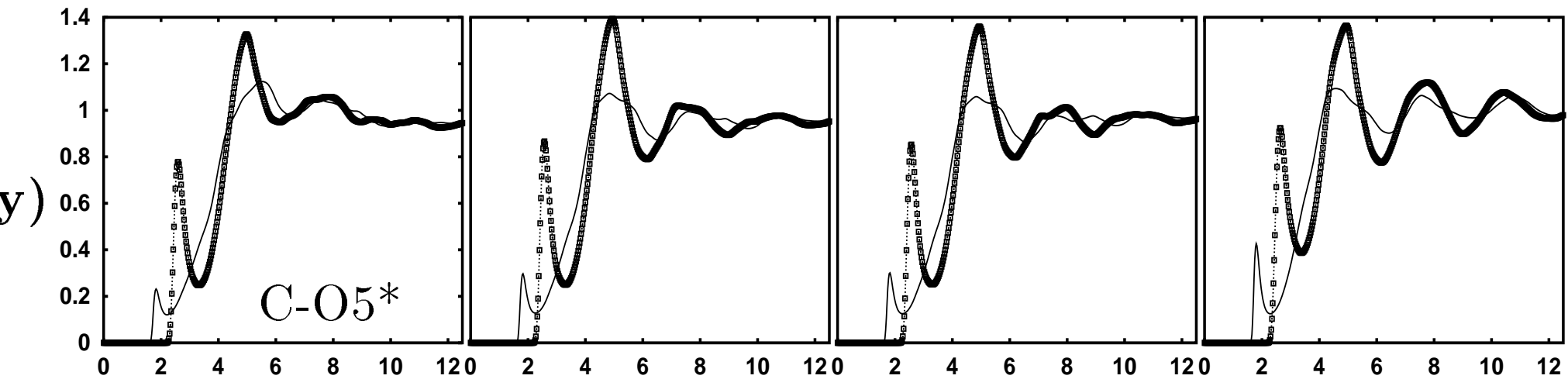
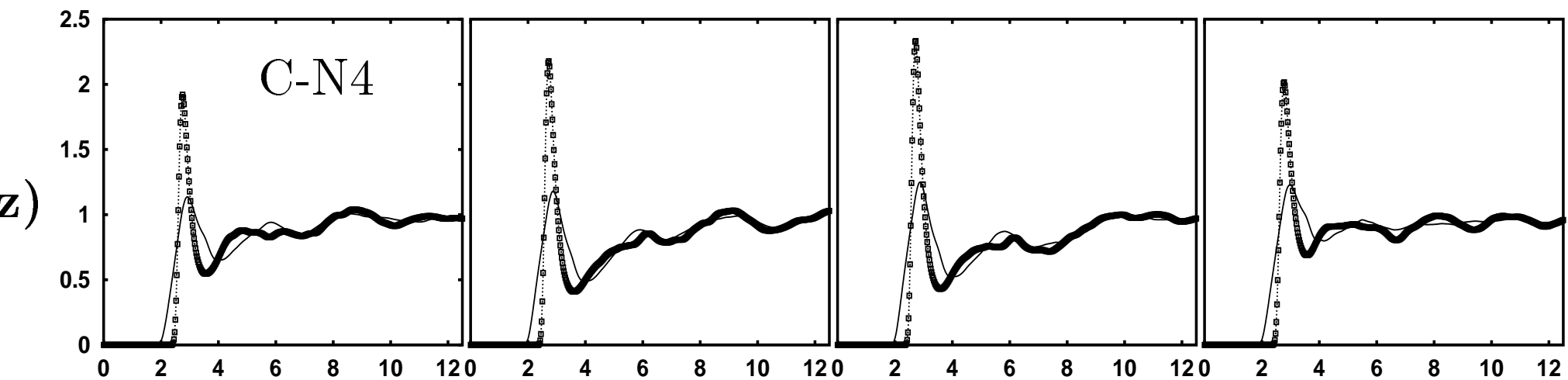
\end{document}